\documentclass[a4paper,12pt]{article}
\usepackage[cp1251]{inputenc}
\usepackage[russian]{babel}

\newcommand{\be}{\begin{equation}}
\newcommand{\ee}{\end{equation}}
\begin{document}

{\Large\bf Local dark energy: HST evidence

from the vicinity of the M 81/M 82 galaxy group}

\vspace{1cm}

{ A.D.~Chernin$^{1,2}$, I.D.~Karachentsev$^3$, O.G.~
Kashibadze$^3$,  D.I.~Makarov$^3$, P.~Teerikorpi$^2$,
M.J.~Valtonen$^{2}$, V.P.~Dolgachev$^1$, L.M.~Domozhilova$^1$ }

\vspace{1cm} {\it $^1$Sternberg Astronomical Institute, Moscow
University,Moscow, 119899, Russia,

$^2$Tuorla Observatory, Turku University, Piikki\"o, 21 500,
Finland,

$^3$Special Astrophysical Observatory, Nizhnii Arkhys, 369167,
Russia}

\vspace{1cm} The Hubble Space Telescope observations of the nearby
galaxy group M 81/M 82 and its vicinity indicate that the
expansion outflow around the group is dominated by the antigravity
of the dark energy background. The local density of dark energy in
the area is estimated to be near the global dark energy density or
perhaps exactly equal to it. This conclusion agrees well with our
previous results for the Local group vicinity and the vicinity of
the Cen A/M 83 group.


\vspace{1cm}

\section{Introduction}

Dark energy was discovered in SN Ia observations at large
cosmological distances, halfway to the cosmic horizon [1,2]. The
discovery was confirmed by the WMAP observations of the cosmic
microwave background isotropy [3,4]. It was found that dark energy
contributes 70-80\% the total energy of the observed Universe
[1-4]. Dark energy drives the accelerated expansion of the
Universe now and for the last 7 Gyr of the cosmic history. The
physical nature and microscopic structure of dark energy are
entirely unknown. Its existence was not predicted by the current
standard model of particle physics; the model cannot explain
either the value of the observed dark energy density. It may only
be assumed that dark energy reflects somehow the unification of
gravity with the other fundamental forces and/or the effects of
quantum physics (see, for instance, a review [5]). In particular,
presently discussed ideas link the origin of dark energy with the
interplay between gravity and electroweak physics in the very
early Universe at the cosmic age of a few picoseconds [6].

Remarkably enough, the macroscopic properties of dark energy are
well known and may be described in a rather complete way. Indeed,
according to the simplest, straightforward and quite likely
interpretation, dark energy is equivalent to the Einstein
cosmological constant $\Lambda$, and it may be considered as a
medium with the constant uniform density:
% 1
\be \rho_V = \Lambda/(8\pi G) > 0.\ee

The dark energy equation of state  has the form [7]:
% 2
\be p_V = - \rho_V.\ee\noindent Here $p_V$ is the dark energy
pressure, $G$ is the gravitational constant and the speed of light
$c = 1$.

The medium with the equation of state given by Eq.(2) is vacuum
[7]: the rest and motion are not discriminated relative to this
medium. The energy density of vacuum is invariant: it is the same
in any reference frame. It was suggested [8] that the
$\Lambda$-vacuum might be identical to the physical vacuum of the
quantum fields; however this idea has not yet been proved or
disproved.

It also follows from Eq.(2) that the vacuum produces antigravity.
According to General Relativity, the effective gravitating density
is determined by both density and pressure of the medium in
combination:
% 3
\be \rho_V + 3 p_V = - 2 \rho_V < 0.\ee \noindent Antigravity is
due to the fact that the effective density of vacuum is negative.
This means that two particles embedded in the vacuum will move
apart each other with acceleration, if they are initially at rest
relative to each other. It is the antigravity of dark energy that
makes the Universe expand with acceleration.

The Hubble magnitude-redshift diagram obtained with the SN Ia
observations [1,2] shows that antigravity of dark energy and the
gravity of matter (baryons and dark matter) balance each other for
a moment at the redshift $z_V \simeq 0.7$. The balance condition
is
% 4
\be \rho_M (z_V) - 2 \rho_V = 0,\ee \noindent where $\rho_M$ is
the matter density. Since the matter density scales with redshift
as $(1 + z)^3$ and the present-day matter density is considered
known, $\rho_M (z=0) \simeq 0.3 \times 10^{-29}$ g cm$^{-3}$, the
estimate of the dark energy density comes from Eq.(4):
% 5
\be \rho_V = \frac{1}{2}\rho_M (z=0) (1 + z_V)^3 \simeq 0.7 \times
10^{-29} \;\; g/cm^{3}.\ee

If dark energy is indeed equivalent to the cosmological constant,
it exists everywhere in space with the same perfectly uniform
density (Eq.5), and all cosmic systems are imbedded in the uniform
dark energy background. We demonstrated [9-20] that the presence
of dark energy on relatively small scales could be recognized
observationally due to its effect on the structure and dynamics of
the Hubble flow of expansion. Dark energy dominates the dynamics
of the expansion not only in the Universe as a whole, but also
within the cosmic cell of uniformity on spatial scales $\le
100-300$ Mpc. Because of this, the flow has a regular structure:
it follows the Hubble linear velocity-distance law, and the
expansion rate (the Hubble constant) is approximately the same on
practically all the spatial scale where the flow is observed.

It is well-known that the expansion flow starts at distances of a
few Mpc from us, in the vicinity of the Local groups of galaxies.
These very local scales of the expansion are especially
interesting. Systematic observations of distances and motions of
galaxies in the Local Group and in the flow around it have been
carried out over the last years with the Hubble Space Telescope
(HST) during more than 200 orbital periods [21-36]. High precision
measurements were made of the radial velocities (with 1-2 km/s
accuracy) and distances (8-10 \% accuracy) for about 200 galaxies
of the Local Group and neighbors from 0 to 7 Mpc from the group
barycenter. We use these data to detect dark energy and estimate
its density at the shortest distances.

In our close galactic vicinity, the gravity is produced by the
matter (dark matter and baryons) of the Local group, and the
antigravity is produced by the local dark energy background. The
effect of dark energy is strong there. The velocity-distance
diagram based on the HST data [21-36] indicates [9-11] that the
local gravity-antigravity balance takes place at a distance $R_V$
which is between 1 and 2 Mpc from the Local group barysenter;
$R_V$ is the radius of the zero-gravity sphere around the Local
group [9-20]. At the smaller distances, the gravity of matter
controls the dynamics of the bound and quasi-stationary group,
while at the larger distances, the antigravity of dark energy
dominates and controls the dynamics of the local expansion flow.

With the known value of the zero-gravity radius $R_V$ and the
measured mass of the Local group $M_{LG} = (1.3 \pm 0.3) \times
10^{12} M_{\odot}$ [22,34], we  estimate the local dark energy
density [12] around the Local group:
% 6
\be \bar \rho_V = \frac{3 M_{LG}}{8 \pi R_V^3} = (0.2-1)\times
10^{-29} \;\; g/cm^3. \ee \noindent The local density of dark
energy proves to be near, if not exactly equal to, the global
figure of Eq.5. This result is completely independent of,
compatible with, and complementary to the largest-distance
observations [1,2]. Note that in both global and local cases, the
estimates have a direct character (contrary to the result [3,4])
and follow essentially the same logic based on the condition of
the gravity-antigravity balance (Eq.4).

The HST data [21-36] may also be used to the study of the
expansion outflows around other galaxy groups in the local volume.
We have recently studied the outflow around the group known as the
Cen A/M83 complex [19] and demonstrated that the dynamical
structure of the flow is significantly affected by the antigravity
of the dark energy background. The local density of dark energy
has been estimated to be near the global cosmological density, or
exactly equal to it, -- in agreement with the result (Eq.6) of the
Local group studies.

In the present paper, we use the HST observations [21-36] and
extend our studies to the nearby M81/M82 galaxy group and its
vicinity. In Sec.2, we give a general discussion of the dark
energy domination effect on the expansion flows within the cosmic
cell of uniformity; in Sec.3, the basic data on the outflow around
the M81/M82 group are presented and analyzed; in Sec.4 the results
are summarized.

\section{Expansion flows dominated by dark energy}

It has long been recognized and recently confirmed [37-48] that
the regular (`cool') flow of expansion is clearly seen from a few
Mpc distance up to the limits of the observed Universe. At all
these distances, the flow preserves its kinematic identity: 1) it
follows the Hubble linear velocity-distance relation, and 2) the
expansion rate (the Hubble factor) is the same -- within the
measurement errors -- for both local and global scales. This is
puzzling because the regular cosmological expansion could only be
expected beyond the scale of the cosmic cell of uniformity, at
distances $\ge 100-300$ Mpc where the matter distribution is
statistically uniform and the Universe is isotropic on average.
How may the flow be regular within the cell of uniformity where
the matter distribution is highly irregular and chaotic?

This problem which is referred to as the Hubble-Sandage paradox
has found its solution with the discovery of dark energy. We argue
[9-15] that the dark energy background with its perfectly uniform
density makes the Universe much more uniform than it might seem
from the matter distribution only. It is important that the dark
energy density is considerably higher than the mean matter density
even inside the cell of matter uniformity, except the areas of
strong matter concentrations like galaxies, galaxy groups or
clusters. Because of the dark energy domination, the Universe may
be described -- in the first and main approximation -- by the
Friedmann isotropic model on all scales larger than, say, 5-10
Mpc. And so on all these scales, one may expect the Hubble flow of
expansion with the Hubble factor:
% 7
\be H = H_V [1 + \rho_M/\rho_V]^{1/2}, \ee \noindent where
% 8
\be H_V = (\frac{8\pi G}{3} \rho_V)^{1/2} = 64 \pm 3 \;\; km/s/Mpc
\ee \noindent is the `universal' expansion rate which is
determined by the dark energy density alone. Here the dark energy
density is taken as $\rho_V = (0.75 \pm 0.07) \times 10^{-29}
g/cm^{3}$ [4]; it is also taken into account that the cosmological
expansion proceeds in the parabolic regime corresponding to the
flat co-moving space [1-4]. At the present epoch, $t = t_0 \simeq
14$ Gyr, the ratio $\rho_M/\rho_V \simeq 1/3$, which leads to
% 9
\be H(t_0) \simeq 74 \pm 6 \;\; km/s/Mpc. \ee\noindent This figure
agrees very well with the most precise current measurements of the
`global' expansion rate reported by the WMAP [4]: $H_G = 74 \pm 4$
km/s/Mpc.

According to Sandage's et al. [42], the expansion rate is $H_0 =
64 \pm 7$ km/s/Mpc in the scale interval from 4 to 200 Mpc. This
observational figure is also rather close to the theory values of
Eqs.8-9. Such a coincidence looks impressive, especially if one
takes into account how different the ways are in which the figures
$H_G, H_0$ and $H(t_0)$ are obtained. The considerations above
suggest that the two observed values for the expansion rate, $H_G$
and $H_0$, are close to each other because each of them is close
to the universal value determined by the dark energy density
alone.

The argument may be inverted: one may consider the regular
expansion flow on the scales 4-200 Mpc as an indication on a
physical factor that acts on all these scales and makes the flow
regular with a common expansion rate, despite the irregularity of
the matter distribution. Then, in the first and main
approximation, one may identify the universal rate $H_V$ with the
observed expansion rate on these scales $H_0$, and find from this
identity the estimate of the local dark energy density:
% 10
\be \bar \rho_V \simeq \frac{3}{8\pi G} H_0^2 \sim 10^{-29} \;\;
g/cm^3.\ee

This estimate shows that the local density is near the global
density of dark energy. The result is completely independent of
the global density measurements [1-4]; it indicates that the dark
energy density may really be the same on both local and global
scales.

Let us turn now to the smallest spatial scales, 1-3 Mpc, on which
the Hubble expansion flow takes its start. It is clear that
cosmology considerations described above cannot be applied
directly to these scales. Indeed, in our close galaxy vicinity,
our Galaxy, the Milky Way, and the Andromeda galaxy move towards
each other, and galaxies around them form a group, the Local
group, in which there is no general expansion. The local expansion
outflow starts in a close vicinity of the group. According to the
HST data [23,34], the `very local' expansion rate $H_L = 72 \pm 6$
km/s/Mpc. Note again a close coincidence of the expansion rates
$H(t_0), H_G, H_0$ and $H_L$ which are found completely
independently of each other.

The local dynamical background of the expansion outflow is
controlled by the gravity of the central group and the antigravity
of the dark energy background in which all the galaxies of the
group and the flow are embedded [9-20]. Considering only the most
important dynamical factors, we may take the gravity field of the
group as nearly centrally-symmetric and static (this is a good
approximation to reality, as exact computer simulations prove
[16-18]. Then, according to the Newtonian gravity law, a galaxy of
the flow is given an acceleration
% 11
\be F_N = - GM/R^2,\ee at its distance $R$ from the group
barycenter (which is the origin of the adopted reference frame).

The local antigravity is produced by the dark energy of vacuum
with the uniform local density $\bar \rho_V$. According to the
`Einstein antigravity law', the dark energy produces acceleration
% 12
\be F_E  = G 2\bar \rho_V (\frac{4 \pi}{3}R^3)/R^2 =  \frac{8
\pi}{3} G \bar \rho_V R, \ee \noindent where $-2 \bar \rho_V$ is
the local effective gravitating density of dark energy (for
details see [9-11]). Eqs.11,12 describe the force field in the
terms of the Newtonian mechanics; a General Relativity equivalent
is given by the static Schwarzschild-de Sitter space-time [11].

It is seen from Eqs.11 and 12 that the gravitational force
($\propto 1/R^2$) dominates over the antigravity force ($\propto
R$) at small distances, and the acceleration is negative there. At
large distances, antigravity dominates, and the acceleration is
positive.  Gravity and antigravity balance each other, so the
acceleration is zero, at the zero-gravity surface which has the
radius
% 13
\be R_V =  (\frac{3}{8\pi}M/\bar \rho_V)^{1/3}. \ee The
zero-gravity surface remains almost unchanged since the formation
of galaxy groups some 10-12 Gyr ago, as the computer simulations
[16-18] indicate.

The zero-gravity radius $R_V$ is a local spatial counterpart of
the `global' redshift $z_V$: the both reflect the
gravity-antigravity balance. However, there is a significant
difference between the global Friedmann theory and the local
theory of Eqs.11-13. Indeed, the global gravity field is uniform
and time-dependent, while the local field is non-uniform and
static. Globally, the gravity-antigravity balance takes place only
at one proper-time moment (at $z = z_V$) in the whole Universe. On
the contrary, the local gravity-antigravity balance exists since
the formation of the central group, but only at one distance ($R =
R_V$).

According to our approach [9-20], the motions of the flow galaxies
originate from the early days of the galaxy group when its major
and minor galaxies participated in violent non-linear dynamics
with multiple collisions and mergers. Our theory and computer
simulations incorporate the concept of the `Little Bang' [49,50]
as a model for the origin of the local expansion flow. The model
shows that some of the dwarf galaxies managed to escape from the
gravitational potential well of the group after having gained
escape velocity from the non-stationary gravity field of the
forming group.

When the escaped galaxies occur beyond the zero-gravity surface
($R > R_V$), their motion is controlled mainly by the dark energy
antigravity and their trajectories are nearly radial there. The
trend of the dynamical evolution controlled by dark energy is seen
from the fact that (as Eqs.11-13 show) at large enough distances
where antigravity dominates over gravity almost completely, the
velocities of the flow are accelerated and finally they grow with
time exponentially: $V \propto \exp [\bar H_V t]$. Because of
this, the expansion flow acquires the linear velocity-distance
relation asymptotically: $V \rightarrow \bar H_V R.$ Here the
value $\bar H_V = (\frac{8 \pi G}{3} \bar \rho_V)^{1/2}$ appears
as the local expansion rate (compare Eq.8) which is constant and
determined by the local dark energy density.

Finally, if the local and global densities of the dark energy are
indeed identical, $\bar \rho_V = \rho_V$, the small-scale
expansion flow around a galaxy group (or a cluster) becomes a
cosmological phenomenon which is consistent with the whole
dynamics of the Universe. In this way, the dark energy antigravity
takes the control over the expansion flows on practically all
spatial scales -- from a few Mpc to the observation horizon -- and
tends to give them the common expansion rate $H_V$ which is
determined by the dark energy density alone.

\section{M 81/M 82 group and the outflow around it}

The M 81/M 82 galaxy group has recently been systematically
observed, alongside with other nearby groups, with the use of the
HST [21-36]. The data on the group are summarized in the Catalog
of Neighboring Galaxies [33] and the recent paper [22]. This is
one the closest groups in the local Universe. Its barycenter is
located at the distance about 3.5 Mpc from the Local group
barycenter. The M 81/M 82 group contains two major galaxies, M 81
and M 82, which have masses $M_{M 81} \simeq 7 \times 10^{11}
M_{\odot}$ and $M_{M 82} \simeq 4 \times 10^{11} M_{\odot}$,
respectively [21]. They are separated by the distance of about 40
kpc and move towards each other with the relative radial velocity
of about 240 km/s. This massive galaxy binary is similar to the
major galaxy binary of the Local group formed by the Milky Way and
the M 31 galaxy; but the M 81 and M 82 galaxies are considerably
closer to each other and consequently their relative velocity is
(two times) larger than that for the Milky Way and the M 31
galaxy. The M 81/M 82 binary is surrounded by a family of smaller
galaxies; 18 of them have precise velocities and distances
measured with the use of the HST [22,33]. The binary and the
galaxies around it form a group which is elongated in shape with
the largest size of about 2 Mpc across. The mean velocity
dispersion in the group is near 70 km/s.

The vicinity of the group up to the 3-4 Mpc from the group
barycenter has also been observed with the use of the HST [22,33].
Around the group, dwarf galaxies are located, and 22 of them have
precise velocities and distances measured with the use of the HST
[22,33]. These outer smaller galaxies move from the group forming
a regular expansion outflow.

Observational data [22,33] on the velocities and distances for the
M 81/M 82 group and the outflow around it are presented in the
Hubble diagram of Fig.1. The velocities and distances are given
relative to the group barycenter. The flow of expansion is clearly
seen in Fig.1 at the distances 1.5-3 Mpc from the group
barycenter. The flow reveals the linear velocity-distance relation
(the Hubble law) with the expansion rate  $H_L = 60 \pm 5$
km/s/Mpc. The flow is cool enough: the velocity dispersion is
about 30 km/s; as this value is affected significantly by the
distance determination errors, the true value is still lower.

Since the major part of the total mass of the group and the flow
is concentrated in the central close binary, the gravity field
produced by the group is practically spherical at distances $\sim
1$ Mpc and more from the group barycenter. Because of this, the
simple relations of Sec.2 may be used as a good approximation for
the description of the dynamical background of the flow. With the
general relation of Eq.13, one may estimate the zero-gravity
radius for the M 81/M 82 group. Taking the mass of the group $M =
M_{M 81} + M_{M 82} \simeq 1 \times 10^{12} M_{\odot}$ and the
dark energy density $\rho_V = 0.75 \times 10^{-29} g/ cm^{3}$, we
find: $R_V \simeq 1$ Mpc.

According to the considerations of the section above, the members
of the galaxy group must be located within the zero-gravity
surface, and this is really obvious from Fig.1. In particular, all
the galaxies with negative velocities are located within the
zero-gravity sphere. The considerations of Sec.2 indicate as well
that the flow of expansion is expected to approach the linear
velocity-distance relation outside the zero-gravity surface where
the antigravity of dark energy is stronger than the gravity of the
group matter. Indeed, Fig.1 shows that the linear
velocity-distance relation emerges from the distance $R \simeq
1.5-2$ Mpc, and all the galaxies at the distances $R
> 1.5$ Mpc move apart of the group.

The theory of Sec.2 makes a definite prediction: the expansion
rate at distances $R > 1.5-2$ Mpc must be near the universal
expansion rate $H_V = 64 \pm 3$ km/s/Mpc, if dark energy has the
same density everywhere in space. The data of Fig.1 agree well
with this prediction, since $H_L \simeq H_V$. On the other hand,
identifying the observed local value $H_L$ with the value $\bar
H_V$ (see Sec.3), one may get an estimate of the local density of
dark energy in the area around the M 81/M 82 group:
% 14
\be \bar \rho_V = (\frac{3}{8 \pi G} H_L^2)^{1/2} = (0.6 \pm
0.2)\times 10^{-29} \;\; g/cm^{-3}.\ee \noindent As we see, the
local density $\bar \rho_V$ is very near the global density
$\rho_V$, if not coinciding with it exactly.

The theory makes also another specific prediction: the velocities
of the local expansion flow must be not less than a minimal
velocity $V_{esc}$:
% 15
\be V_{esc} = (\frac{2G M}{R_V})^{1/2} (\frac{R}{R_V})^{1/2} [1 +
\frac{1}{2} (\frac{R}{R_V})^3 - \frac{3}{2}
(\frac{R}{R_V})]^{1/2}. \ee

The minimal velocity corresponds to the minimal total mechanical
energy,
% 16
\be E_{esc} = - \frac{3}{2} \frac{GM}{R_V}, \ee \noindent needed
for a particle to escape from the gravitational potential well of
the group. Due to the antigravity of dark energy, this energy is
negative. The condition $V \ge V_{esc}$ is comfortably satisfied,
if one assumes that $\bar \rho_V = \rho_V$ and also adopts
somewhat larger value for the total mass of the group, namely $M =
M_{M 81} + M_{M 82} = 1.3 \times 10^{12} M_{\odot}$ (which is
within the limits of its observational determination). Then the
zero-gravity radius $R_V = 1.3$ Mpc. With this mass and the
assumed dark energy density, all the galaxies of the flow are
located above the critical line $V_{esc}(R)$ showed by a bold
curve in Fig.1. The critical velocity is zero at the distance $R =
R_V = 1.3$ Mpc.

On the other hand, a clear structure of the Hubble diagram of
Fig.1 enables to recognize independently the position of the
zero-gravity sphere in the diagram. Indeed, since the zero-gravity
surface lies outside the group volume, it should be that $R_V > 1
$ Mpc. The fact that the linear velocity-distance relation is seen
from a distance of about, say, 2 Mpc suggests that $R_V < 2$ Mpc.
If so, Eq.13 leads directly to the robust upper (from $R > 2$ Mpc)
and lower ($R < 3$ Mpc) limits of the local dark energy density:
% 17
\be 0.2  < \bar \rho_V < 1 \times 10^{-29} \;\; g/cm^3. \ee
\noindent This result is in good agreement with the estimate of
Eq.14 and the considerations above. The lower limit in Eq.17 is
most significant: it means that the dark energy does exist in the
area. In combination, both limits imply that the value of the
local dark energy density is near the value of the global dark
energy density, or may be exactly equal to it.

One may see from Fig.1 that the structure of the flow follows the
trend of the minimal velocity: the linear regression line of the
flow (the thin line in Fig.1) is nearly parallel to the minimal
velocity curve, at $R > R_V$. It may easily be seen from the
theory of Sec.2 that in the limit of large distances, the minimal
velocity and the real velocity of the flow galaxies have a common
asymptotic, $V = V_{esc} = H_V R$, independently on the initial
conditions of the galaxy motions.

For a comparison, a minimal escape velocity, $(\frac{2G
M}{R_V})^{1/2} (\frac{R}{R_V})^{1/2}$, is showed in Fig.1 for a
`no-vacuum model' with zero dark energy density (dashed line). The
real flow is obviously ignores the trend of the minimal velocity
in this case: the velocities of the flow grow with distance, while
the minimal velocity decreases. It is seen also that two galaxies
of the flow at $R > R_V$ violate obviously the no-vacuum model:
they are located below the dashed line, in the diagram. This
comparison is clearly in favor of the vacuum model and against the
model with no dark energy.

\section{Conclusions}

As is well-known, galaxy groups and their close environment are
more or less similar to each others in the local Universe [51-54].
Similar conclusion follows also from large N-body $\Lambda$CDM
cosmological simulations [55-60]. The structure with a massive
galaxy group (or cluster) in its center and a cool expansion
outflow outside -- a `Hubble cell' [19] -- seems to be
characteristic for the local Universe.

High accuracy observations of the local volume with the use of the
HST [21-36] enabled us to study the Local Group and the cool
expansion outflow around it as an archetypical example of the
structure. The major new physics here is the presence of dark
energy and its domination in the dynamics of the expansion flows.
The basic physical quantity of the cell is the zero-gravity radius
$R_V$ introduced in [9-13]. For the Local Hubble cell, $R_V = 1.25
$ Mpc. The central group of the cell is located within the
zero-gravity surface ($R < R_V$) and controlled mostly by the
gravity of the group matter (dark matter and baryons). The cool
expansion outflow develops outside the surface ($R > R_V$), and
its structure and evolution are determined mainly by the
antigravity of the dark energy background. The HST observations
[22,33] enable us also to study a similar dynamical structure in
the area of the Cen A/M 83 galaxy group [19]; the third clear
example of the structure is demonstrated above in the present
paper.

Our observations and analysis of the three nearby Hubble cells
lead to the following conclusions: 1) dark energy is definitely
present in the local Universe on the spatial scales of a few Mpc;
2) the antigravity of dark energy dominates the dynamics of the
expansion outflows on these scales; 3) the local density of dark
energy is near the global density of dark energy or perhaps
exactly equal to it. The conclusions form a strong evidence for
dark energy as cosmic vacuum which is equivalent to the Einstein
cosmological constant.

The work of A.C., V.D. and L.D. was partly supported by a RFBR
grant 06-02-16366.

\section*{References}

\noindent [1] A.G. Riess, A.V. Filippenko, P. Challis et al. AJ,
{\bf }, 1009 (1998).

\noindent [2] S. Perlmuter, G. Aldering, G. Goldhaber G. et al.
ApJ, {\bf 517}, 565 (1999).

\noindent [3] D.N. Spergel et al. ApJS, {\bf 148}, 175 (2003).

\noindent [4] D.N. Spergel et al. astro-ph/0603449 (2006).

\noindent [5] S. Weinberg. Rev. Mod. Phys., {\bf 61} 1 (1989).

\noindent [6] N. Arkani-Hamed et al. Phys. Rev. Lett., {\bf 85}
4434 (2000).

\noindent [7] E.B. Gliner. Sov.Phys. JETP, {\bf 22}, 378 (1966).

\noindent [8] Ya.B. Zeldovich. Sov. Phys.-Uspekhi {\bf 95} 209
(1968).

\noindent [9] A.D. Chernin. Physics-Uspekhi, {\bf 44}, 1099
(2001).

\noindent [10]  A.D. Chernin, P. Teerikorpi, Yu.V. Baryshev. Adv.
Space Res., {\bf 31}, 459, (2003).

\noindent [11] A.D. Chernin, P. Teerikorpi, Yu.V. Baryshev. A\& A,
{\bf 456}, 13 (2006).

\noindent [12] A.D. Chernin, I.D. Karachentsev, P. Teerikorpi et
al. (2007) (in press).

\noindent [13] I.D. Karachentsev, A.D. Chernin, P. Teerikorpi.
Astrofizika {\bf 46}, 491 (2003).

\noindent [14] Yu. Baryshev, A. Chernin, P. Teerikorpi. A\&A, {\bf
378}, 729 (2001).

\noindent [15] P. Teerikorpi, A.D. Chernin, Yu.V. Baryshev. A\&A,
{\bf 440}, 791 (2005).

\noindent [16] A.D. Chernin, I.D. Karachentsev, M.J. Valtonen et
al. A\&A, {\bf 467}, 933 (2005).

\noindent [17] A.D. Chernin, I.D. Karachentsev, M.J. Valtonen et
al. A\&A, {\bf 415}, 19 (2004).

\noindent [18] V.P. Dolgachev, L.M. Domozhilova, A.D. Chernin.
Astr. Rep., {\bf 47}, 728 (2003).

\noindent [19] A.D. Chernin, I.D. Karachentsev, D.I. Makarov et
al. astro-ph/0704.2753 (2007).

\noindent [20] A.D. Chernin, I.D. Karachentsev, P. Teerikorpi et
al. astro-ph/0706.4068 (2007).

\noindent [21] I.D. Karachentsev et al. A\&A, {\bf 389}, 812
(2002).

\noindent [22] I.D. Karachentsev, O.G. Kashibadze. Astrofizika
{\bf 49}, 5 (2006).

\noindent [23] I.D. Karachentsev et al. astro-ph/0603091 (2006).

\noindent [24] I.D. Karachentsev, M.E. Sharina, E.K. Grebel et al.
A\&A, {\bf 352}, 399 (1999).

\noindent [25] A.E. Dolphin, L.M. Makarova, I.D. Karachentsev et
al. MNRAS, {\bf 324}, 249 (2001).

\noindent [26] I.D. Karachentsev, M.E. Sharina, A.E. Dolphin et
al. A\&A, {\bf 379}, 407 (2001).

\noindent [27] I.D. Karachentsev, A.E. Dolphin, D. Geisler et al.
A\&A, {\bf 383}, 125 (2002).

\noindent [28]  I.D. Karachentsev, M.E. Sharina, A.E. Dolphin et
al. A\&A, {\bf 385}, 21 (2001).

\noindent [29] I.D. Karachentsev, M.E. Sharina, D.I. Makarov et
al. A\&A, {\bf 389}, 812 (2002).

\noindent [30] I.D. Karachentsev, D.I. Makarov, M.E. Sharina et
al. A\&A, {\bf 398}, 479 (2003).

\noindent [31] I.D. Karachentsev, E.K. Grebel, M.E. Sharina et al.
A\&A, {\bf 404}, 93 (2003).

\noindent [32]  I.D. Karachentsev, M.E. Sharina, A.E. Dolphin,
E.K. Grebel. A\&A, {\bf 408}, 111 (2003).

\noindent [33] I.D. Karachentsev, V.E. Karachentseva, W.K.
Huchtmeier, D.I. Makarov. AJ, {\bf 127}, 2031 (2004).

\noindent [34] I.D. Karachentsev. AJ, {\bf 129}, 178 (2005).

\noindent [35] I.D. Karachentsev, A.E. Dolphin, R.B. Tully. AJ,
{\bf 131}, 1361 (2006).

\noindent [36] I.D. Karachentsev, R.B. Tully, A.E. Dolphin et al.
AJ, {\bf 133}, 504 (2007).

\noindent [37] A. Sandage. ApJ, {\bf 307}, 1 (1986).

\noindent [38] A. Sandage. ApJ, {\bf 527}, 479 (1999).

\noindent [39] A. Sandage et al. ApJ, {\bf 172}, 253 (1972).

\noindent [40] F. Thim, G. Tammann, A. Saha et al. ApJ, {\bf 590},
256 (2003).

\noindent [41] A. Sandage, G.A. Tamman, B. Reindl. A\&A, {\bf
424}, 43 (2004).

\noindent [42] A. Sandage, G.A. Tamman, A. Saha et al. ApJ, {\bf
653}, 843 (2006).

\noindent [43] R. Rekola, M.G. Richer, M.L. McCall et al. MNRAS,
361, 330 (2005).

\noindent [44] T. Ekholm, P. Teerikorpi, G. Theureau et al. A\&A,
{\bf 347}, 99 (1999).

\noindent [45] P. Teerikorpi, M. Hanski, G. Theureau. et al. A\&A,
{\bf 334}, 395 (1998).

\noindent [46] G. Theureau, M. Hanski, T. Ekholm et al.  A\&A,
{\bf 322}, 730 (1997).

\noindent [47] T. Ekholm, Yu. Baryshev, P. Teerikorpi et al. A\&A,
{\bf 368}, 17 (2001).

\noindent [48] G. Paturel, P. Teerikorpi. A\&A, {\bf 443}, 883
(2005)

\noindent [49] G.G. Byrd, M.J. Valtonen, M. McCall, K. Innanen.
AJ, {\bf 107}, 2055 (1994).

\noindent [50] M.J. Valtonen, M. McCall, K. Innanen,  J.-Q. Zheng,
G.G.Byrd (1995) in Dark Matter, AIP Conf. Proc. 336, p.450 (eds.
S.S. Holt and C.L. Bennett).

\noindent [51] S. van den Bergh. astro-ph/0305042 (2003).

\noindent [52] S. van den Bergh. AJ, {\bf 124}, 782 (2002).

\noindent [53] S van den Bergh. ApJ, {\bf 559}, L113 (2001).

\noindent [54]  M.J. Valtonen, G.G. Byrd. ApJ, {\bf 303}, 523
(1986).

\noindent [55] K. Nagamine, R. Cen, J.P. Ostriker. Bul. Amer.
Astron. Soc., {\bf 31}, 1393 (1999).

\noindent [56] K. Nagamine, J.P. Ostriker, R. Cen. ApJ. {\bf 553},
513 (1991).

\noindent [57] J.P. Ostriker, Y. Suto. ApJ, {\bf 348}, 378 (1990).

\noindent [58] Y. Suto, R. Cen, J.P. Ostriker. ApJ, {\bf 395}, 1
(1992).

\noindent [59]  V.A. Strauss, R. Cen, J.P. Ostriker. ApJ, {\bf
408}, 389 (1993).

\noindent [60] A.V. Macci\`{o}, F. Governato, G. Horellou. MNRAS,
{\bf 359}, 941 (2005).

\section*{Figure caption}

Fig.1. The Hubble velocity-distance diagram for the M 81/M 82
galaxy group and its vicinity, according to the HST observations
[22,33]. The velocities and distances are given relative to the
group barycenter. The galaxies of the group are located within the
zero-gravity sphere of the radius $R_V = 1.3$2 Mpc. The flow of
expansion starts in the outskirts of the group; all the galaxies
at distances $R > R_V$ Mpc move apart of the group (positive
velocities). The flow reveals the linear velocity-distance
relation (see also the text).

\end{document}